\documentclass[twocolumn,showpacs,preprintnumbers,amsmath,amssymb,prb]{revtex4} 

\usepackage{graphicx}
\usepackage{dcolumn}
\usepackage{bm}
\usepackage{amsmath}

\begin{document}


\title{Time-resolved charge detection with cross-correlation techniques}


\author{B. K\"ung}
\email{kuengb@phys.ethz.ch}
\affiliation
{Solid State Physics Laboratory, ETH Zurich,
8093 Zurich, Switzerland}

\author{O. Pf\"affli}
\affiliation
{Solid State Physics Laboratory, ETH Zurich,
8093 Zurich, Switzerland}

\author{S. Gustavsson}
\affiliation
{Solid State Physics Laboratory, ETH Zurich,
8093 Zurich, Switzerland}

\author{T. Ihn}
\affiliation
{Solid State Physics Laboratory, ETH Zurich,
8093 Zurich, Switzerland}

\author{K. Ensslin}
\affiliation
{Solid State Physics Laboratory, ETH Zurich,
8093 Zurich, Switzerland}

\author{M. Reinwald}
\affiliation{Institut f\"ur Experimentelle und Angewandte Physik, Universit\"at Regensburg, 93040 Regensburg, Germany}

\author{W. Wegscheider}
\affiliation{Institut f\"ur Experimentelle und Angewandte Physik, Universit\"at Regensburg, 93040 Regensburg, Germany}


\date{January 13, 2009}

\begin{abstract}
We present time-resolved charge-sensing measurements on a GaAs double quantum dot with two proximal quantum
point-contact (QPC) detectors. The QPC currents are analyzed with cross-correlation techniques, which enable us to
measure dot charging and discharging rates for significantly smaller signal-to-noise ratios than required for charge
detection with a single QPC. This allows us to reduce the current level in the detector and therefore the invasiveness
of the detection process and may help to increase the available measurement bandwidth in noise-limited setups.
\end{abstract}

\pacs{73.23.Hk, 73.40.Gk}

\maketitle


The use of quantum point contacts (QPCs) as charge sensors integrated in semiconductor quantum dot (QD)
structures\cite{Field93} has become a well-established experimental technique in current nanoelectronics research. The
time-resolved operation of such sensors\cite{Schleser04, Vandersypen04, Fujisawa04b} allows us to observe the charge
and spin dynamics of single electrons\cite{Elzerman04, Gustavsson06a} which has potential applications in
metrology\cite{Gustavsson08a} or for the implementation of qubit readout schemes in quantum information
processing.\cite{Petta05} Another appealing property of the QD-QPC system is that it opens the possibility of studying
a well-defined quantum mechanical measurement process and testing the theory of measurement-induced
decoherence.\cite{Buks98}

The difficulty in achieving quantum-limited charge detection is mainly the limited bandwidth of the readout circuit
compared to charge coherence times. In addition, decoherence mechanisms exist that are due to the QPC but not directly
linked to detection, such as the excitation of electrons in the QD driven by noise in the QPC,\cite{Gustavsson07a} an
effect which is more pronounced at higher source-drain voltages. Both problems are related to the limit in
signal-to-noise ratio (SNR) offered by present-day setups. A common experimental approach to overcome such a limitation
is the use of cross correlation of independent measurement channels. In the context of charge sensing, correlation
techniques have previously been used in Al single electron transistor setups to suppress background charge
noise\cite{Buehler03} and to obtain estimates for the spatial distribution of sources thereof.\cite{Zorin96}
High-frequency noise measurements usually rely on correlation techniques which eliminates noise contributions of the
wiring and the amplifiers.\cite{Reznikov95}

In the present work we present cross-correlated charge sensing measurements in a double quantum dot (DQD) sample with
two charge readout QPCs. The potential advantages of such a design for the continuous quantum measurement of charge
qubit oscillations have been put forward by Jordan and B\"uttiker.\cite{Jordan05} While the corresponding time scales
are yet beyond our experimentally achievable bandwidth, we demonstrate the benefit of cross-correlation techniques in
the classical detection of electron tunneling. By a detailed analysis of the cross-correlation function of the QPC
currents and of higher-order correlators, we are able to measure tunneling rates in a manner eliminating uncorrelated
amplifier noise. Compared to a measurement of the same quantities using only one channel, we are able to reduce the
detector current by roughly 1 order of magnitude.

\begin{figure}[b]
\includegraphics{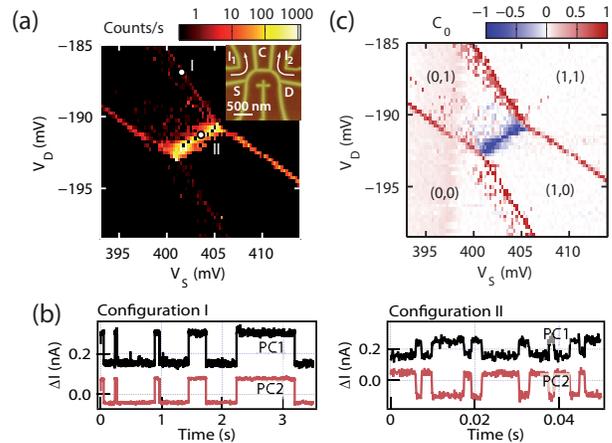}
\caption{(Color online) (a) Inset: AFM micrograph of the sample which consists of two QDs in series (QD1 and QD2) with
two charge readout QPCs, denoted PC1 and PC2. The source, drain, and center barriers can be tuned with in-plane gates
S, D, and C. Main graph: part of the DQD charge stability diagram obtained by counting the number of switching events
in $I_2$. (b) Detector currents recorded at two different gate configurations indicated in (a). Dot-lead tunneling
processes (I) cause identical switching directions in both QPCs whereas inter-dot processes (II) cause opposite
switching. (c) Current correlator $C_0 = \langle I_1 I_2 \rangle/(\langle I_1^2 \rangle \langle I_2^2 \rangle)^{1/2}$
extracted from the raw data used in (a), revealing the correlation-anticorrelation pattern in the $V_S$-$V_D$ plane.
Numbers $(n,m)$ indicate the electron occupancy of the dots relative to the state $(0,0)$. }
\label{fig:Corr_figure_Intro}
\end{figure}

The inset of Fig.~\ref{fig:Corr_figure_Intro}(a) shows the structure, fabricated on a GaAs/AlGaAs heterostructure
containing a two-dimensional electron gas $34 \, \mathrm{nm}$ below the surface (density: $5 \times 10^{15} \,
\mathrm{m}^{-2}$ and mobility: $40 \, \mathrm{m}^2/\mathrm{Vs}$ at $4.2 \, \mathrm{K}$). The electron gas was locally
depleted by anodic oxidation with an atomic force microscope (AFM).\cite{Fuhrer02} The measurements were performed in a
$^3\mathrm{He}/^4\mathrm{He}$ dilution refrigerator with an electron temperature of about $80\, \mathrm{mK}$, as
determined from the width of thermally broadened Coulomb blockade resonances.\cite{Kouwenhoven97} The structure
consists of two QDs in series (denoted QD1 and QD2) with two charge-readout QPCs (PC1 and PC2). The strength of the
tunneling coupling to source and drain leads is tuned with the gates denoted S and D; gate C controls the interdot
coupling and is kept at a constant voltage for these measurements.

Both QPCs are voltage biased and tuned to conductances below $2e^2/h$. Their currents are measured with an $I/V$
converter with a bandwidth of $19 \, \mathrm{kHz}$ and sampled at a rate of $50 \, \mathrm{kS/s}$. The data is stored
for further processing in the form of time traces typically few seconds long. Electrons entering or leaving either dot
cause steps in the currents that can be counted.\cite{Schleser04} Figure \ref{fig:Corr_figure_Intro}(a) shows a color
plot of the count rate in PC2 vs S and D gate voltages close to a pair of triple points of the DQD
system\cite{Vanderwiel03} at zero source-drain voltage. Lines with negative slope belong to equilibrium tunneling
events between the dots and the leads. The inter-dot charging energy ($0.3\, \mathrm{meV}$) is much larger than the
thermal energy, therefore also the line of inter-dot tunneling events with positive slope is observable. The
corresponding tunneling rate of about $1\, \mathrm{kHz}$ is the largest in the system. Few additional counts outside
the main resonances are due to excitation processes driven by the currents in the QPCs
(Ref.~\onlinecite{Gustavsson07a}) (source-drain voltage $300\, \mathrm{\mu V}$).

Due to geometry, the capacitive coupling between the QPCs and the QDs is asymmetric; charging QD1 will, for example,
cause a larger step in the conductance of PC1 than charging QD2. Accordingly, the steps due to dot-lead tunneling
events have the same sign in both QPCs whereas inter-dot events cause opposite switching as seen in the time traces
plotted in Fig.~\ref{fig:Corr_figure_Intro}(b). A simple parameter which characterizes the correlation between the two
channels is the correlator
\begin{equation}
\label{eq:Corr_correlator} C_0 = \frac{\langle I_1 I_2 \rangle -
\langle I_1 \rangle \langle I_2 \rangle  } {\sqrt{\langle I_1^2
\rangle - \langle I^{\vphantom{1}}_1 \rangle^2} \sqrt{\langle
I_2^2 \rangle - \langle I^{\vphantom{1}}_2 \rangle^2}},
\end{equation}
where angular brackets denote time averaging. We obtain this quantity, as well as any other cross-correlation
expression discussed later in this paper, by digital processing of the raw time trace data. In
Fig.~\ref{fig:Corr_figure_Intro}(c), we plot $C_0$ calculated from the same data as used in panel (a). It clearly
displays the expected pattern of positive and negative correlations along the charging lines of the DQD stability
diagram. Note that in the following, we implicitly assume the mean values of $I_1$ and $I_2$ to be subtracted by
setting $\langle I_1 \rangle = \langle I_2 \rangle = 0$.

Going beyond this more qualitative information, in the following we analyze how to extract physical tunneling rates
with the help of cross-correlation techniques and apply this to the example of tunneling from the lead into and out of
QD2 (rates $\Gamma_\mathrm{in}$ and $\Gamma_\mathrm{out}$) in the present sample. The underlying problem is to extract
these two characteristic parameters of a random telegraph signal (RTS) $I^{(c)}$ which is, as we assume, a component of
both QPC currents, along with uncorrelated noise. If the noise is stronger than the signal, the information on the
actual time dependence of $I^{(c)}$ is lost even if there are two measurement channels available. This is however not a
problem since one can determine the rates $\Gamma_\mathrm{in,out}$ entirely on the basis of \emph{time-averaged}
quantities derived from $I^{(c)}$. For the analysis presented here, these are on the one hand its autocorrelation
function from which we can extract a characteristic time constant $\tau_0 = 1/(\Gamma_\mathrm{in} +
\Gamma_\mathrm{out})$ and on the other hand its skewness $\gamma$ which depends on the occupation probabilities of the
high and low current states of $I^{(c)}$ and allows one to determine the ratio $\Gamma_\mathrm{in} /
\Gamma_\mathrm{out}$. The sought-after $\Gamma_\mathrm{in,out}$ are then uniquely determined by $\tau_0$ and $\gamma$.
This concept of exploiting third-order cumulants of a telegraph signal for measurement has also been discussed in
Ref.~\onlinecite{Jordan05prb}.

\begin{figure}
\includegraphics{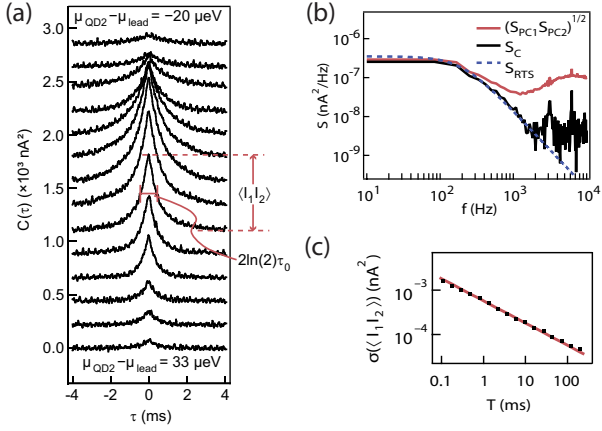}
\caption{(Color online) (a) Evolution of the current cross-correlation function $C(\tau) = \langle
I_{1}(t+\tau)I_{2}(t) \rangle$ when crossing the $(1,0) \rightarrow (1,1)$ charging line in the DQD stability diagram
[outside the scan range of Fig.~\ref{fig:Corr_figure_Intro}(b)]. The curves are offset for clarity; going from the
lowest to the uppermost curve corresponds to adding one electron to QD2. $C(\tau)$ exhibits an exponential decay $\sim
\exp(-|\tau|/\tau_0)$ characteristic for the random telegraph signal present in both QPC currents. The center curves,
which feature the largest amplitude $C(0)$, belong to QPC signals with rather balanced occupation of the high- and
low-current states. (b) Noise reduction due to cross correlation in time traces measured on a Coulomb peak. Plotted are
the Fourier transform $S_C$ of the cross-correlation function and the geometric mean of the two QPC current spectra
$S_\mathrm{PC1}$ and $S_\mathrm{PC2}$. The dashed line represents the ideal RTS spectrum $S_\mathrm{RTS} (f) = 2
\langle I_\mathrm{RTS}^2 \rangle \tau_0 /[1+\tau_0^2 (2\pi f)^2]$ with parameters $\langle I_\mathrm{RTS}^2 \rangle =
0.22 \times 10^{-3} \, \mathrm{nA}^2$ and $\tau_0 = 0.8\, \mathrm{ms}$. (c) Standard deviation $\sigma (\langle I_1 I_2
\rangle )$ of the average $\langle I_1 I_2 \rangle = T^{-1}\int_0^T I_1(t) I_2(t) \mathrm{d} t$ as a function of the
integration time $T$. The solid line marks the expected $\sim T^{-1/2} $ behavior. }
\label{fig:Corr_figure_Spectra_XCorrEvo}
\end{figure}

To state this more precisely, we split up the QPC currents according to $I_j = \alpha_j I^{(c)} + I^{(n)}_j$, $j = 1,
\, 2$, where $\alpha_j$ are dimensionless factors ($\alpha_1 > 0$ by convention) and $I_j^{(n)}$ are mutually
uncorrelated noise components. The product of $I_1$ and $I_2$ appearing in the cross-correlation function $C (\tau) =
\langle I_1(t) I_2(t+\tau) \rangle $ then consists of four terms among which any one containing a factor $I^{(n)}_1$ or
$I^{(n)}_2$ is integrated to zero. The only nonvanishing part is then proportional to the autocorrelation function of
the signal $I^{(c)}$,
\begin{eqnarray}
\label{eq:Corr_XCorrFunction}
C(\tau) & \approx & \alpha_1 \alpha_2 \langle I^{(c)}(t) I^{(c)}(t+\tau) \rangle  \nonumber\\
& = & \alpha_1 \alpha_2 \langle I^{(c)2} \rangle e^{-|\tau |/\tau_0},
\end{eqnarray}
where the decay time of the exponential is given by $\tau_0 = 1/( \Gamma_\mathrm{in} +
\Gamma_\mathrm{out})$.\cite{Machlup54} Note that form (\ref{eq:Corr_XCorrFunction}) of $C(\tau)$ implies a purely
Poissonian tunneling process. On time scales relevant for our measurements, non-Poissonian statistics can occur when
excited dot states are involved\cite{Gustavsson06b} and would manifest itself in a deviation of $C(\tau)$ from the
exponential shape. Figure \ref{fig:Corr_figure_Spectra_XCorrEvo}(a) shows a set of $C(\tau)$ curves belonging to the
crossover from the (1,0) to the (1,1) state in the DQD charge stability diagram. For curves in the center of this plot,
the electrochemical potential of QD2 is roughly aligned with that of the lead, and the tunneling in and out rates are
similar. The peak amplitude of $C(\tau)$ is largest in this regime. It is proportional to $\langle I^\mathrm{(c)2}
\rangle$ which is maximum in the case of a symmetric RTS, as we discuss later in more detail. Moving away from this
point, the peak amplitude decays. The behavior of the peak width outside the resonance is determined by the behavior of
the rates $\Gamma_\mathrm{in,out}$: While one of the rates tends to zero, the other approaches its finite saturation
value which is also the saturation value of $1/\tau_0$. The peak width therefore remains nonzero.

The noise reduction due to the cross correlation is best visualized in the frequency domain. In
Fig.~\ref{fig:Corr_figure_Spectra_XCorrEvo}(b), we plot the geometric mean of the power spectral densities of some
example time traces $I_1$ and $I_2$ along with the Fourier transform of their cross-correlation function. The spectrum
of the raw traces consists of the Lorentzian contribution of the telegraph signal and a noise background on the order
of $10^{-7} \, \mathrm{nA^2/Hz}$ which is dominated by the (current-independent) noise of the room-temperature $I/V$
converter and contains an additional current-dependent part that is most likely related to charge noise in the sample.
In the cross-correlation spectrum, the signal part is unchanged; the noise on the other hand is clearly suppressed.
This remains, for the moment, a qualitative statement, and we postpone the quantitative discussion about the noise
reduction to the end of the paper.

The correlation time $\tau_0$ gives the sum of the two tunneling rates but is insensitive to their relative magnitude.
A second experimental parameter is therefore needed which depends on $\Gamma_\mathrm{in}/\Gamma_\mathrm{out}$ and is
also accessible in high-noise conditions. It is natural to consider the skewness $\gamma = \langle I^{(c)3} \rangle /
\langle I^{(c)2} \rangle ^{3/2}$ because it parametrizes the degree of asymmetry in the current distribution function
of $I^{(c)}$, which is in turn fully determined by $\Gamma_\mathrm{in}/\Gamma_\mathrm{out}$. Namely, the occupation
probability of the low-current state of the RTS (electron on the dot) is $p_\mathrm{low} =
\Gamma_\mathrm{in}/(\Gamma_\mathrm{in} + \Gamma_\mathrm{out})$; analogously $p_\mathrm{high} =
\Gamma_\mathrm{out}/(\Gamma_\mathrm{in} + \Gamma_\mathrm{out})$. Assuming a current difference of $\Delta I$ between
the two states and $\langle I^{(c)} \rangle = 0$, then the $n$th central moment of $I^{(c)}$ is given by
\begin{eqnarray}
\label{eq:Corr_central_moments}
\langle I^{(c)n} \rangle = p_\mathrm{low} (-p_\mathrm{high}\Delta I)^n + p_\mathrm{high} (p_\mathrm{low}\Delta I)^n \nonumber\\
=\frac{\Gamma_\mathrm{in} \Gamma_\mathrm{out}}{(\Gamma_\mathrm{in} + \Gamma_\mathrm{out})^{n+1}} [\Gamma_\mathrm{in}^{n-1}-(-\Gamma_\mathrm{out})^{n-1}]\Delta I^n.
\end{eqnarray}
In calculating the skewness based on Eq.~(\ref{eq:Corr_central_moments}) for $n=2$ and 3, we see that the current scale
$\Delta I$, i.e., the information on the strength of the QD-QPC coupling, is eliminated. After some algebra, we obtain
the expression
\begin{eqnarray}
\label{eq:Corr_skewness} \gamma = \frac{\langle I^{(c)3} \rangle}{\langle I^{(c)2} \rangle^{3/2}} = \frac{
\Gamma_\mathrm{in}-\Gamma_\mathrm{out}}{(\Gamma_\mathrm{in} \Gamma_\mathrm{out})^{1/2}}.
\end{eqnarray}
Using Eq.~(\ref{eq:Corr_skewness}) and the previously determined $\tau_0 = 1/(\Gamma_\mathrm{in} + \Gamma_\mathrm{out})$, we can now write down the total event rate,
\begin{eqnarray}
\Gamma_\mathrm{tot} &=& \frac{\Gamma_\mathrm{in} \Gamma_\mathrm{out}}{\Gamma_\mathrm{in} +\Gamma_\mathrm{out}} \nonumber\\
&=& (\Gamma_\mathrm{in} +\Gamma_\mathrm{out})  \frac{\Gamma_\mathrm{in} \Gamma_\mathrm{out}}{\Gamma_\mathrm{in}^2 + 2\Gamma_\mathrm{in} \Gamma_\mathrm{out} +\Gamma_\mathrm{out}^2} \nonumber\\
&=& \frac{1}{\tau_0(4+\gamma^2)}.
\end{eqnarray}
The individual tunneling rates are then given by
\begin{eqnarray}
\label{eq:Corr_both_rates} \Gamma_{\mathrm{in/out}} = \frac{2}{\tau_0 (4+\gamma^2 \mp \gamma \sqrt{4+\gamma^2} )}.
\end{eqnarray}

The skewness is experimentally accessed through an appropriate combination of second- and third-order correlators
computed from the raw time traces $I_1$ and $I_2$ that have the property to be insensitive to the background noise. On
the one hand, we use again $\langle I_1 I_2 \rangle$ which is the cross-correlation function at zero time difference
$C(\tau = 0)$ and is equal to $\alpha_i \alpha_j \langle I^{(c)2} \rangle$. On the other hand, we use the combinations
$ \langle I_i^2 I^{\vphantom{1}}_j \rangle$ which are proportional to the third moment of $I^{(c)}$. In writing the QPC
currents as a sum of signal and noise, $I_j = \alpha_j I^{(c)} + I^{(n)}_j$, it is readily seen that any term
containing the noise $I_j^{(n)}$ gives zero contribution to the time average and we have $ \langle I_i^2
I^{\vphantom{1}}_j \rangle \approx \alpha_i^2 \alpha^{\vphantom{1}}_j \langle I^{(c)3} \rangle$. The skewness can then
be expressed as
\begin{eqnarray}
\label{eq:Corr_gamma_experimentally} \gamma \approx \mathrm{sgn} \left( \langle I^{\vphantom{1}}_1 I_2^2 \rangle
\right) \left( \frac{\langle I_1^2 I^{\vphantom{1}}_2 \rangle \langle I^{\vphantom{1}}_1 I_2^2 \rangle}{\langle I_1 I_2
\rangle^3} \right)^{1/2}.
\end{eqnarray}
The asymmetry in this formula is caused by our previous choice $\alpha_1 > 0$; i.e., we fixed the sign of $I^{(c)}$
such that it is positively correlated with $I_1$. This freedom of choice is not unique to our correlation analysis.
Instead, the assignment of one detector event type (e.g., ``PC1 current up'') to one system event type (e.g.,
``electron tunneling from QD2 into lead'') has to be done in any case.

\begin{figure}
\includegraphics{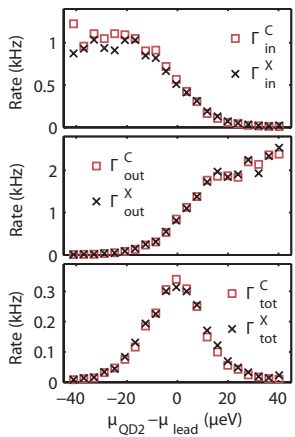}
\caption{(Color online) Plots of the rates $\Gamma_\mathrm{in}$, $\Gamma_\mathrm{out}$, and
$\Gamma_\mathrm{tot}=\Gamma_\mathrm{in}\Gamma_\mathrm{out}/(\Gamma_\mathrm{in}+\Gamma_\mathrm{out})$ as determined by
electron counting (marked with the letter ``C'') and by current cross-correlation (``X''). The sweep range is indicated
by a line in the top graph of Fig.~\ref{fig:Corr_figure_Comparison}(a).} \label{fig:Corr_figure_Rates_1d}
\end{figure}

Before turning to the experimental results, we discuss the role of the integration time $T$ in the cross-correlation
process. How large do we have to choose $T$ until the cross-correlation function (\ref{eq:Corr_XCorrFunction}) is
reproduced to a good accuracy? In order to estimate the remaining noise contribution to $C(\tau)$ after averaging, we
treat the integration as a summation over samples that are separated in time by the typical autocorrelation time of the
noise $\tau_{n}$ and are therefore statistically independent. Using the central limit theorem, we write the standard
deviation of this sum as $(\tau_{n} / T )^{1/2} ( \sigma_1^{(n)} \sigma_2^{(n)} )^{1/2}$
[cf.~Fig.~\ref{fig:Corr_figure_Spectra_XCorrEvo}(a), inset], where we have introduced the symbols $\sigma_j^{(n)} =
\langle I^{(n)2}_j \rangle^{1/2}$ for the noise in the channels. It should not exceed the contribution of the telegraph
signal $\alpha_1 \alpha_2 \langle I^{(c)2} \rangle $. For the data presented here, time traces were recorded for $5 \,
\mathrm{s}$ and digitally low-pass filtered at $3 \, \mathrm{kHz}$ ($\tau_{n} \approx 0.1 \, \mathrm{ms}$), yielding an
expected noise reduction of 0.005. As seen from Eq.~(\ref{eq:Corr_central_moments}), the quantity $\alpha_1 \alpha_2
\langle I^{(c)2} \rangle $ contains a factor $\Gamma_\mathrm{in}\Gamma_\mathrm{out}$ and is therefore small whenever
one of the tunneling rates is small. As a result, the situation where the two rates are similar presents the optimal
case for a correlation measurement.

Even disregarding any uncorrelated noise, the exponential shape of the autocorrelation function of $I^{(c)}$ is the
limit of infinite integration time. It is practically reached under the condition that $T$ covers a sufficient number
of switching events, $T \gg 1/\Gamma_\mathrm{tot}$. This second condition on $T$ is therefore linked to the statistical
uncertainty of the measurement.

In Fig.~\ref{fig:Corr_figure_Rates_1d}, we compare the outcome of the conventional (counting) method and the
correlation procedure for a constant bias of $222 \, \mathrm{\mu V}$ across the QPCs. The two data sets are generally
in good agreement, with small systematic deviations on the sides of the Coulomb peak and a certain scatter due to low
statistics in the tails. The observed asymmetry between tunneling in and out processes (i.e., the difference in the
maximum values of $\Gamma_\mathrm{in}$ and $\Gamma_\mathrm{out}$) can be explained by the existence of a second
degenerate quantum state in QD2.

\begin{figure}
\includegraphics{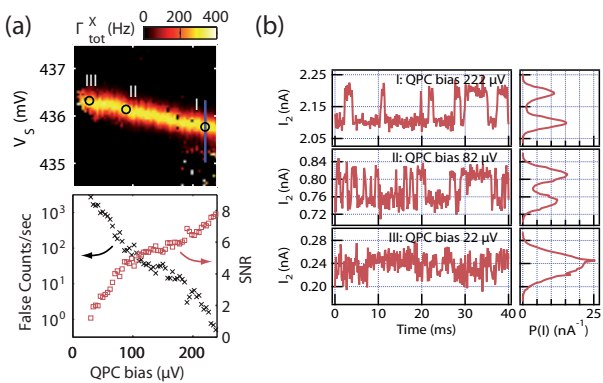}
\caption{(Color online) (a) Top: tunneling rate $\Gamma^X_\mathrm{tot}$ between QD2 and lead as a function of QPC bias
(applied to both PC1 and PC2) and S gate voltage determined by cross-correlation analysis. Bottom: SNR for the counting
analysis and expected false count rate as a function of QPC bias. The false count rate was calculated from the SNR data
assuming Gaussian amplifier noise with an autocorrelation time of $\tau_{n}=0.1\, \mathrm{ms}$. A SNR of 6 will result
in a false count rate on the order of $10 \, \mathrm{Hz}$ and can therefore be considered as the minimum requirement
for the counting analysis. It is reached for bias voltages above approximately $150 \, \mathrm{\mu V}$. (b) Examples of
the time dependence (left column) and current distribution function $P(I)$ (right column) of $I_2$ recorded at three
different QPC bias voltages indicated in the top graph in (a). The RTS as a component of the current is recognized by
the naked eye in all three cases, but only the trace ``I'' allows for a determination of the transition rates without
significant error when analyzed with a counting algorithm. } \label{fig:Corr_figure_Comparison}
\end{figure}

Having checked the consistency of the two methods in a regime where both are applicable, we test the correlation method
in a regime with smaller signal levels. We do this by reducing the source-drain voltage on the QPCs. The step height
$\alpha_j \Delta I^{(c)}$ of the RTS is approximately proportional to the bias whereas the noise level $\sigma_j^{(n)}$
remains constant. The ratio of the two is the SNR relevant for the standard counting analysis. an insufficient SNR will
result in systematic measurement errors due to false counts, namely, an overestimation of the slower rate in case of an
asymmetric RTS, or of both rates in case of a symmetric RTS. Assuming a certain current distribution of the amplifier
noise around the discrete current levels of the RTS, say a Gaussian distribution, the false count rate can be estimated
as the number of statistically independent current measurements that lie outside a distance $\alpha_j \Delta I^{(c)}/2$
from the mean. We can express it with the help of the error function as $0.5 [1-\mathrm{erf(SNR}/2\sqrt{2})]/\tau_{n}$.
The lower plot in Fig.~\ref{fig:Corr_figure_Comparison}(a) shows a measurement of the signal-to-noise ratio along with
the estimated false count rate calculated in this manner. The value for the SNR considered sufficient depends on the
desired accuracy; here we require a SNR of more than 6 which results in a false count rate on the order of $10 \,
\mathrm{Hz}$ and which is reached for source-drain voltages larger than $150 \, \mathrm{\mu V}$.

In comparison, the measurement of $\Gamma_\mathrm{tot}$ shown in the upper plot of
Fig.~\ref{fig:Corr_figure_Comparison}(a) demonstrates that the cross-correlation analysis is applicable down to
significantly lower bias voltages, therefore reducing both the power dissipated by the sensors and the energy scale of
the emitted radiation. As discussed, the best results are obtained close to the maximum of the peak where the rate is
measured reliably, i.e., with fluctuations below the statistical uncertainty due to the finite number of detected
events, down to bias voltages of $22 \, \mathrm{\mu V}$. Only below (and in the tails of the peak) the errors grow and
eventually the analysis algorithm fails.

We now formulate a more precise criterion for comparing the two methods. In particular, it is first of all necessary to
quantify the residual noise. For this purpose, we define $\sigma_X^{(n)}$ as the standard deviation of the fluctuations
in the function $C(\tau)$ [cf.~Fig.~\ref{fig:Corr_figure_Spectra_XCorrEvo}(a)] measured in the absence of a RTS signal.
The ratio $(\sigma_X^{(n)} / \sigma_1^{(n)} \sigma_2^{(n)})^{1/2}$ can be considered as a measure for the success in
suppressing the noise by current cross correlation. However, the quantitative meaning of the noise level in the
correlation case is different compared to the counting case. The actual parameter of interest is the measurement
uncertainty caused by this noise. Calculating it in the general case is a nontrivial task, on the one hand, because of
the complexity of the analysis algorithm and, on the other hand, because of the many experimental variables that play a
role such as the absolute value of $\Gamma_\mathrm{tot}$, RTS asymmetry, measurement bandwidth, noise spectrum, and
differences between the two channels (i.e., in the parameters $\alpha^{\vphantom{n}}_j$ and $\sigma_j^{(n)}$). We
therefore restrict our discussion to the specific measurement situation discussed in this paper, in particular, to the
case of nearly identically coupled QPCs ($\alpha_1 = \alpha_2 = 1$). We ask this question: by how much, starting from
the limiting counting SNR of 6, can we reduce the signal strength $\Delta I$ until we expect the correlation procedure
to generate the same absolute error of about $10 \, \mathrm{Hz}$ in $\Gamma_\mathrm{tot}$? We write this ``figure of
merit'' as
\begin{eqnarray}
\label{eq:Corr_fom_factorized} \frac{\Delta I_{\mathrm{min},X}}{\Delta I_{\mathrm{min},C}} &=& \frac{\Delta
I_{\mathrm{min},X}}{\sqrt{\sigma^{(n)}_X}} \frac{\sqrt{\sigma^{(n)}_X}}{\sqrt{\sigma^{(n)}_1 \sigma^{(n)}_2}}
\frac{\sqrt{\sigma^{(n)}_1 \sigma^{(n)}_2}}{\Delta I_{\mathrm{min},C}}.
\end{eqnarray}
The third factor in Eq.~(\ref{eq:Corr_fom_factorized}) is the original (inverse) SNR for the counting algorithm. The
first factor can be considered as the analog for the cross-correlation case, relating the signal strength to the
residual noise $\sigma_X^{(n)}$ in $C(\tau)$. It was determined with a numerical simulation. In applying the data
analysis algorithm to randomly generated time traces imitating the experimental ones (symmetric RTS with overlaid
Gaussian noise, low-pass filtering with $3 \, \mathrm{kHz}$, $\Gamma_\mathrm{tot} = 0.3 \, \mathrm{kHz}$, and $T = 5 \,
\mathrm{s}$), the measurement uncertainty is obtained from the scatter in the output. The minimum $\Delta I$ for an
error below $10 \, \mathrm{Hz}$ determined in this way was given by $11 \, (\sigma^{(n)}_X)^{1/2}$. Finally, the second
factor in Eq.~(\ref{eq:Corr_fom_factorized}) is the noise reduction achieved in experiment; we measured $\sigma^{(n)}_X
\approx 2.2 \times 10^{-24} \, \mathrm{A}^2$, $\sigma^{(n)}_1 \approx 21 \times 10^{-12} \, \mathrm{A}$, and
$\sigma^{(n)}_2 \approx 16 \times 10^{-12} \, \mathrm{A}$. Plugging in these numbers we find
\begin{eqnarray}
\label{eq:Corr_fom_numerically} \frac{\Delta I_{\mathrm{min},X}}{\Delta I_{\mathrm{min},C}} &\approx & 11 \cdot
\frac{1}{\sqrt{150}} \cdot \frac{1}{6} = 0.15.
\end{eqnarray}
This means that in the case of the correlation experiment one can obtain meaningful values for the tunneling rates for signal-to-noise ratios approaching 1.

To summarize, we have measured charge fluctuations on a GaAs DQD in a time-resolved manner simultaneously with two QPC
charge sensors. By evaluating their cross-correlation function and third-order correlators, we are able to determine
the two time constants of tunneling back and forth between one dot and the adjacent lead. Obtaining the same
information directly from either of the two QPC signals requires a significantly larger RTS amplitude because of the
limitation due to amplifier noise. An interesting prospect is the application of the correlation technique to
radio-frequency QPC setups\cite{Mueller07, Reilly07, Cassidy07} where it would allow us to push the shot-noise
limitation to the detection bandwidth toward the regime of charge qubit coherence times.

The authors thank Yuval Gefen and Lieven Vandersypen for fruitful discussion. Financial support from the Swiss National Science Foundation
(Schweizerischer Nationalfonds) is gratefully acknowledged.



\end{document}